\documentclass[12pt]{article}
\usepackage[table]{xcolor}
\usepackage{colortbl}
\definecolor{lightgray}{gray}{0.9}
\usepackage{graphicx}
\usepackage{lscape}
\usepackage{citesort}
\usepackage{amssymb}
\usepackage{appendix}
\usepackage{multirow}

\usepackage{graphicx}
\usepackage{lscape}
\usepackage{citesort}
\usepackage{amssymb}
\usepackage{appendix}
\usepackage{multirow}

\usepackage{mathrsfs}
\begin{document}
\def\qq{\langle \bar q q \rangle}
\def\uu{\langle \bar u u \rangle}
\def\dd{\langle \bar d d \rangle}
\def\sp{\langle \bar s s \rangle}
\def\GG{\langle g_s^2 G^2 \rangle}
\def\Tr{\mbox{Tr}}
\def\figt#1#2#3{
        \begin{figure}
        $\left. \right.$
        \vspace*{-2cm}
        \begin{center}
        \includegraphics[width=10cm]{#1}
        \end{center}
        \vspace*{-0.2cm}
        \caption{#3}
        \label{#2}
        \end{figure}
    }

\def\figb#1#2#3{
        \begin{figure}
        $\left. \right.$
        \vspace*{-1cm}
        \begin{center}
        \includegraphics[width=10cm]{#1}
        \end{center}
        \vspace*{-0.2cm}
        \caption{#3}
        \label{#2}
        \end{figure}
                }

\def\ds{\displaystyle}
\def\beq{\begin{equation}}
\def\eeq{\end{equation}}
\def\bea{\begin{eqnarray}}
\def\eea{\end{eqnarray}}
\def\beeq{\begin{eqnarray}}
\def\eeeq{\end{eqnarray}}
\def\ve{\vert}
\def\vel{\left|}
\def\ver{\right|}
\def\nnb{\nonumber}
\def\ga{\left(}
\def\dr{\right)}
\def\aga{\left\{}
\def\adr{\right\}}
\def\lla{\left<}
\def\rra{\right>}
\def\rar{\rightarrow}
\def\lrar{\leftrightarrow}
\def\nnb{\nonumber}
\def\la{\langle}
\def\ra{\rangle}
\def\ba{\begin{array}}
\def\ea{\end{array}}
\def\tr{\mbox{Tr}}
\def\ssp{{\Sigma^{*+}}}
\def\sso{{\Sigma^{*0}}}
\def\ssm{{\Sigma^{*-}}}
\def\xis0{{\Xi^{*0}}}
\def\xism{{\Xi^{*-}}}
\def\qs{\la \bar s s \ra}
\def\qu{\la \bar u u \ra}
\def\qd{\la \bar d d \ra}
\def\qq{\la \bar q q \ra}
\def\gGgG{\la g^2 G^2 \ra}
\def\q{\gamma_5 \not\!q}
\def\x{\gamma_5 \not\!x}
\def\g5{\gamma_5}
\def\sb{S_Q^{cf}}
\def\sd{S_d^{be}}
\def\su{S_u^{ad}}
\def\sbp{{S}_Q^{'cf}}
\def\sdp{{S}_d^{'be}}
\def\sup{{S}_u^{'ad}}
\def\ssp{{S}_s^{'??}}

\def\sig{\sigma_{\mu \nu} \gamma_5 p^\mu q^\nu}
\def\fo{f_0(\frac{s_0}{M^2})}
\def\ffi{f_1(\frac{s_0}{M^2})}
\def\fii{f_2(\frac{s_0}{M^2})}
\def\O{{\cal O}}
\def\sl{{\Sigma^0 \Lambda}}
\def\es{\!\!\! &=& \!\!\!}
\def\ap{\!\!\! &\approx& \!\!\!}
\def\md{\!\!\!\! &\mid& \!\!\!\!}
\def\ar{&+& \!\!\!}
\def\ek{&-& \!\!\!}
\def\kek{\!\!\!&-& \!\!\!}
\def\cp{&\times& \!\!\!}
\def\se{\!\!\! &\simeq& \!\!\!}
\def\eqv{&\equiv& \!\!\!}
\def\kpm{&\pm& \!\!\!}
\def\kmp{&\mp& \!\!\!}
\def\mcdot{\!\cdot\!}
\def\erar{&\rightarrow&}
\def\olra{\stackrel{\leftrightarrow}}
\def\ola{\stackrel{\leftarrow}}
\def\ora{\stackrel{\rightarrow}}

\def\simlt{\stackrel{<}{{}_\sim}}
\def\simgt{\stackrel{>}{{}_\sim}}


\title{
         {\Large
                 {\bf
                    On the  strong coupling $N^{(*)}N^{(*)}\pi$
                 }
         }
      }

\author{\vspace{1cm}\\
{\small K. Azizi$^a$ \thanks {e-mail: kazizi@dogus.edu.tr}\,, Y.
Sarac$^b$
\thanks {e-mail: yasemin.sarac@atilim.edu.tr}\,\,, H.
Sundu$^c$ \thanks {e-mail: hayriye.sundu@kocaeli.edu.tr}} \\
{\small $^a$  Department of Physics, Do\u gu\c s University, Ac{\i}badem-Kad{\i}k\"oy, 34722 Istanbul, Turkey} \\
{\small $^b$ Electrical and Electronics Engineering Department,
Atilim University, 06836 Ankara, Turkey} \\
{\small $^c$ Department of Physics, Kocaeli University, 41380 Izmit,
Turkey}}
\date{}

\begin{titlepage}
\maketitle
\thispagestyle{empty}

\begin{abstract}
 We study the strong vertices $N^*N\pi$, $N^*N^*\pi$ and $NN\pi$ in QCD, where $N^*$ denotes the negative parity $N (1535)$ state. We use  the most
 general form of the interpolating currents  to calculate the corresponding strong coupling constants. It is obtained that the coupling associated to $N^*N\pi$ vertex is strongly suppressed compared to
 those related to two other vertices. The strong coupling corresponding to  $N^*N^*\pi$ is obtained to be roughly half of that of  $NN\pi$ vertex.  We compare the
 obtained results  on $N^*N\pi$ and $NN\pi$ vertices with the existing predictions of  other theoretical studies as well as  those
extracted from the experimental data.

\end{abstract}

~~~PACS number(s):14.20.Dh, 14.40.Be, 13.75.Gx, 11.55.Hx
\end{titlepage}

\section{Introduction}
In hadron physics the strong couplings among various particles play  essential roles in
understanding of the strong interaction and the nature and structure
of the participating particles. Hence, one needs the precise determination
of these strong coupling constants. Especially, due to the limited experimental information
related to the decay properties of the negative parity baryons, the
theoretical studies in this area can provide effective contributions.
Beside the properties of negative parity baryons, reinvestigation of
the coupling form factor of the pion-nucleon and comparisons of the
results with the known theoretical and empirical results may supply a
better understanding about the nonperturbative nature of QCD.

The properties of negative parity nucleon, such as mass and the
other spectroscopic properties, were studied extensively
(see for instance the Ref.~\cite{Chung,Lee,Kondo,Oka1,Oka} and the
references therein). Beside the spectral properties, its magnetic
moment was analyzed using QCD sum rules~\cite{Aliev} and effective
Hamiltonian \cite{Narodetskii} approaches. The light cone QCD sum
rule was applied to obtain the electromagnetic transition
form factors of $\gamma^*N\rightarrow N(1535)$  and $\gamma^\ast N\rightarrow N^\ast(1520)$ ~\cite{Aliev1,Aliev2}.  The radiative transition of negative to positive parity
nucleon was also studied in Ref.~\cite{Azizi3}. This
work is devoted to the study of the strong coupling constants among the  negative  parity $N^*(1535)$,
nucleon  and pion in QCD. Such type of
investigations were done extensively using either three point QCD
sum rules with Ioffe current or  light cone QCD sum rules and some
other  methods. One can find some of them in
Refs.~\cite{Meissner,Meissner1,Aliev3,Morimatsu,Shiomi,Birse,Birse1,Jido1,Hosaka,Oka6,Kim,Kim1,Zhu,Kim2,Kim3,Ericson,Downum,Ericson1}
and the references therein. In the present study, regarding the mass
difference between the initial and final nucleons in the considered
transitions, the three point QCD sum rule method is applied. This method~\cite{Shifman} is one of the most powerful
method among  nonperturbative approaches and was used extensively
and successfully to obtain the hadronic properties. We consider  the
three-point correlation function  with the most general form of the
interpolating currents for the positive and negative parity nucleons.

The outline of the article is as follows. In section 2 the details
of the QCD sum rules calculation for the considered transitions are
presented. Section 3 is devoted to the numerical results and
conclusion.

\section{The strong coupling form factors between the positive and negative parity nucleons and pion}
This section presents some details of the calculations of the
strong coupling form factors  between the positive and negative parity nucleons and pion using the QCD sum rule method. The starting point is to consider the following correlation function
 in terms of the interpolating fields of the
considered states:
\begin{eqnarray}\label{CorrelationFunction}
\Pi(q)=i^2 \int d^4x~ \int d^4y~e^{-ip\cdot x}~
e^{ip^{\prime}\cdot y}~{\langle}0| {\cal T}\left (
J_{N^{('*)}}(y)~J_{\pi}(0)~\bar{J}_{N^{('*)}}(x)\right)|0{\rangle},
\end{eqnarray}
where ${\cal T}$ is the time ordering operator and  $q=p-p'$ is the
transferred momentum. In this equation  $J_i$
represent the interpolating fields of nucleon and $\pi$ meson. $N$
and $N^*$ denote the positive parity ground state and the negative parity $N (1535)$, respectively. Here, $N'$  stands for the excited positive parity state $N (1440)$, that also couples to the nucleon currents and we shall
take into account its contribution to the correlation function. We calculate this correlation function in terms of physical and OPE (operator product expansion) sides. By matching
 these two representations, the corresponding strong coupling form factors are found. To suppress the contributions of the higher states and continuum,
a double Borel transformation with respect to $p^2$ and $p'^2$ together with continuum subtractions are applied to both sides.

\subsection{Physical Side}

For the calculation of the physical side  complete sets of
appropriate $N$, $N^*$, $N^{'}$ and $\pi$ hadronic states carrying the same
quantum numbers with the corresponding interpolating currents are
placed into the correlation function.  Integrations over $x$ and
$y$  gives
\begin{eqnarray} \label{physicalside}
\Pi^{Phy}(q)&=&\frac{\langle 0 \mid
 J_{N}\mid N(p^{\prime},s^{\prime})\rangle \langle 0 \mid
 J_{\pi}\mid \pi(q)\rangle  \langle N(p^{\prime},s^{\prime})\pi(q)\mid
N(p,s)\rangle\langle N(p,s)\mid
 \bar{J}_{N}\mid 0\rangle }{(p^2-m_{N}^2)(p^{\prime^2}-m_{N}^2)(q^2-m_{\pi}^2)}
 \nonumber \\
&+&\frac{\langle 0 \mid
 J_{N^*}\mid N^*(p^{\prime},s^{\prime})\rangle \langle 0 \mid
 J_{\pi}\mid \pi(q)\rangle  \langle N^*(p^{\prime},s^{\prime})\pi(q)\mid
N^*(p,s)\rangle\langle N^*(p,s)\mid
 \bar{J}_{N^*}\mid 0\rangle }{(p^2-m_{N^*}^2)(p^{\prime^2}-m_{N^*}^2)(q^2-m_{\pi}^2)}
 \nonumber \\
 &+&\frac{\langle 0 \mid
 J_{N}\mid N(p^{\prime},s^{\prime})\rangle \langle 0 \mid
 J_{\pi}\mid \pi(q)\rangle  \langle N(p^{\prime},s^{\prime})\pi(q)\mid
N^*(p,s)\rangle\langle N^*(p,s)\mid
 \bar{J}_{N^*}\mid 0\rangle }{(p^2-m_{N^*}^2)(p^{\prime^2}-m_{N}^2)(q^2-m_{\pi}^2)}
 \nonumber \\
&+& \frac{\langle 0 \mid
 J_{N^*}\mid N^*(p^{\prime},s^{\prime})\rangle \langle 0 \mid
 J_{\pi}\mid \pi(q)\rangle  \langle N^*(p^{\prime},s^{\prime})\pi(q)\mid
N(p,s)\rangle\langle N(p,s)\mid
 \bar{J}_{N}\mid 0\rangle }{(p^2-m_{N}^2)(p^{\prime^2}-m_{N^*}^2)(q^2-m_{\pi}^2)}
\nonumber \\
&+& \frac{\langle 0 \mid
 J_{N^{\prime}}\mid N^{\prime}(p^{\prime},s^{\prime})\rangle \langle 0 \mid
 J_{\pi}\mid \pi(q)\rangle  \langle N^{\prime}(p^{\prime},s^{\prime})\pi(q)\mid
N^{\prime}(p,s)\rangle\langle N^{\prime}(p,s)\mid
 \bar{J}_{N^{\prime}}\mid 0\rangle }{(p^2-m_{N^{\prime}}^2)(p^{\prime^2}
 -m_{N^{\prime}}^2)(q^2-m_{\pi}^2)}
\nonumber \\
&+& \frac{\langle 0 \mid
 J_{N}\mid N(p^{\prime},s^{\prime})\rangle \langle 0 \mid
 J_{\pi}\mid \pi(q)\rangle  \langle N(p^{\prime},s^{\prime})\pi(q)\mid
N^{\prime}(p,s)\rangle\langle N^{\prime}(p,s)\mid
 \bar{J}_{N^{\prime}}\mid 0\rangle }{(p^2-m_{N^{\prime}}^2)(p^{\prime^2}
 -m_{N}^2)(q^2-m_{\pi}^2)}
\nonumber \\
&+& \frac{\langle 0 \mid
 J_{N^{\prime}}\mid N^{\prime}(p^{\prime},s^{\prime})\rangle \langle 0 \mid
 J_{\pi}\mid \pi(q)\rangle  \langle N^{\prime}(p^{\prime},s^{\prime})\pi(q)\mid
N(p,s)\rangle\langle N(p,s)\mid
 \bar{J}_{N}\mid 0\rangle }{(p^2-m_{N}^2)(p^{\prime^2}
 -m_{N^{\prime}}^2)(q^2-m_{\pi}^2)}
\nonumber \\
&+& \frac{\langle 0 \mid
 J_{N^{*}}\mid N^{*}(p^{\prime},s^{\prime})\rangle \langle 0 \mid
 J_{\pi}\mid \pi(q)\rangle  \langle N^{*}(p^{\prime},s^{\prime})\pi(q)\mid
N^{\prime}(p,s)\rangle\langle N^{\prime}(p,s)\mid
 \bar{J}_{N^{\prime}}\mid 0\rangle }{(p^2-m_{N^{\prime}}^2)(p^{\prime^2}
 -m_{N^{*}}^2)(q^2-m_{\pi}^2)}
\nonumber \\
&+& \frac{\langle 0 \mid
 J_{N^{\prime}}\mid N^{\prime}(p^{\prime},s^{\prime})\rangle \langle 0 \mid
 J_{\pi}\mid \pi(q)\rangle  \langle N^{\prime}(p^{\prime},s^{\prime})\pi(q)\mid
N^{*}(p,s)\rangle\langle N^{*}(p,s)\mid
 \bar{J}_{N^{*}}\mid 0\rangle }{(p^2-m_{N^{*}}^2)(p^{\prime^2}
 -m_{N^{\prime}}^2)(q^2-m_{\pi}^2)}
\nonumber \\
&+& \cdots~.
\end{eqnarray}
where we included contributions of both positive and negative  parity
nucleons and the contributions coming from the higher states and
continuum are represented by $\cdots$. The matrix elements emerging
in this result are parameterized  in terms of  the residues
$\lambda_{N}$,   $\lambda_{N^*}$ and $\lambda_{N^{'}}$,  the spinors  $u_{N}$,  $u_{N^*}$ and $u_{N^{'}}$ ,
the leptonic decay constant of $\pi$ meson as well as the  strong couplings
 $g_{NN\pi}$, $g_{N^*N^*\pi}$,  $g_{N^{'}N^{'}\pi}$, $g_{NN^*\pi}$, $g_{N^*N\pi}$,  $g_{NN^{'}\pi}$, $g_{N^{'}N\pi}$, $g_{N^{'}N^*\pi}$ and  $g_{N^*N^{'}\pi}$ as
\begin{eqnarray}\label{matriselement}
\langle 0 \mid
 J_{N^{(\prime)}}\mid N^{(\prime)}(p^{\prime},s^{\prime})\rangle&=&\lambda_{N^{(')}}
 u_{N^{(\prime)}}(p^{\prime},s^{\prime}),
\nonumber \\
\langle 0 \mid
 J_{N^*}\mid N^*(p^{\prime},s^{\prime})\rangle&=&\lambda_{N^*} \gamma_5 u_{N^*}(p^{\prime},s^{\prime}),
 \nonumber \\
\langle 0 \mid
 J_{\pi}\mid
\pi (q)\rangle&=&i\frac{f_{\pi}m_{\pi}^{2}}{m_u+m_d},
\nonumber \\
\langle  N^{(\prime)}(p^{\prime},s^{\prime})\pi(q) \mid
N^{(\prime)}(p,s)\rangle&=&g_{N^{(\prime)}N^{(\prime)}\pi}\bar{u}_{N^{(\prime)}}
(p^{\prime},s^{\prime})i\gamma_{5}u_{N^{(\prime)}}(p,s),
\nonumber \\
\langle  N^*(p^{\prime},s^{\prime})\pi(q) \mid
N^*(p,s)\rangle&=&g_{N^*N^*\pi}\bar{u}_{N^*}(p^{\prime},s^{\prime})i\gamma_{5}u_{N^*}(p,s),
\nonumber \\
\langle  N^*(p^{\prime},s^{\prime})\pi(q) \mid
N^{(\prime)}(p,s)\rangle&=&g_{N^{(\prime)}N^*\pi}\bar{u}_{N^*}(p^{\prime},s^{\prime})i
u_{N^{(\prime)}}(p,s),
\nonumber \\
\langle  N^{(\prime)}(p^{\prime},s^{\prime})\pi(q) \mid
N^*(p,s)\rangle&=&g_{N^*N^{(\prime)}\pi}\bar{u}_{N^{(\prime)}}(p^{\prime},s^{\prime})i
u_{N^*}(p,s).
\end{eqnarray}
 Using these parametrizations and summation over Dirac
spinors via
\begin{eqnarray}\label{spinorsum}
\sum_{s} u_N(p,s)\bar{u}_N(p,s)=\not\!p+m_N
\end{eqnarray}
 in Eq.(\ref{physicalside}), we get
\begin{eqnarray} \label{physicalside1}
\Pi^{Phy}(q)&=&i^2\frac{f_{\pi}m_{\pi}^2}{m_u+m_d}\Bigg\{
g_{NN\pi}
\lambda_N^2\frac{(\not\!p^{\prime}+m_N)\gamma_5(\not\!p+m_N)}
{(p^2-m_N^2)(p^{\prime^2}-m_{N}^2)(q^2-m_{\pi}^2)}
\nonumber \\
&-&g_{N^*N^*\pi}
\lambda_{N^*}^2\frac{(\not\!p^{\prime}-m_{N^*})\gamma_5(\not\!p-m_{N^*})}
{(p^2-m_{N^*}^2)(p^{\prime^2}-m_{N^{*}}^2)(q^2-m_{\pi}^2)}
\nonumber \\
&+&g_{N^*N\pi}\lambda_{N}
\lambda_{N^*}\frac{(\not\!p^{\prime}+m_N)\gamma_5(\not\!p-m_{N^*})}
{(p^2-m_{N^*}^2)(p^{\prime^2}-m_{N}^2)(q^2-m_{\pi}^2)}
\nonumber \\
&-&g_{NN^*\pi}\lambda_{N}
\lambda_{N^*}\frac{(\not\!p^{\prime}-m_{N^*})\gamma_5(\not\!p+m_N)}
{(p^2-m_{N}^2)(p^{\prime^2}-m_{N^*}^2)(q^2-m_{\pi}^2)}
\nonumber \\
&+& g_{N^{\prime}N^{\prime}\pi}
\lambda_{N^\prime}^2\frac{(\not\!p^{\prime}+m_{N^{\prime}})\gamma_5(\not\!p+m_{N^{\prime}})}
{(p^2-m_{N^{\prime}}^2)(p^{\prime^2}-m_{N^{\prime}}^2)(q^2-m_{\pi}^2)}
\nonumber \\
&+& g_{N^{\prime}N\pi} \lambda_N
\lambda_{N^\prime}\frac{(\not\!p^{\prime}+m_{N})\gamma_5(\not\!p+m_{N^{\prime}})}
{(p^2-m_{N^{\prime}}^2)(p^{\prime^2}-m_{N}^2)(q^2-m_{\pi}^2)}
\nonumber \\
&+& g_{NN^{\prime}\pi} \lambda_N
\lambda_{N^\prime}\frac{(\not\!p^{\prime}+m_{N^{\prime}})\gamma_5(\not\!p+m_{N})}
{(p^2-m_{N}^2)(p^{\prime^2}-m_{N^{\prime}}^2)(q^2-m_{\pi}^2)}
\nonumber \\
&-&g_{N^{\prime}N^*\pi}\lambda_{N^{\prime}}
\lambda_{N^*}\frac{(\not\!p^{\prime}-m_{N^*})\gamma_5(\not\!p+m_{N^{\prime}})}
{(p^2-m_{N^{\prime}}^2)(p^{\prime^2}-m_{N^*}^2)(q^2-m_{\pi}^2)}
\nonumber \\
&+&g_{N^*N^{\prime}\pi}\lambda_{N^{\prime}}
\lambda_{N^*}\frac{(\not\!p^{\prime}+m_{N^{\prime}})\gamma_5(\not\!p-m_{N^*})}
{(p^2-m_{N^{*}}^2)(p^{\prime^2}-m_{N^{\prime}}^2)(q^2-m_{\pi}^2)}
 \Bigg\}.
\end{eqnarray}

After application of a double Borel transformation with
respect to the initial and final momenta the final result for the physical
side in terms of different Dirac structures becomes
\begin{eqnarray} \label{Borelphysicalside}
\widehat{\textbf{B}}\Pi^{Phys}(q)&=&i^2\frac{f_{\pi}m_{\pi}^2}{(m_u+m_d)(q^2-m_{\pi}^2)}\Big(
\Phi_1\not\!p\not\!q\gamma_5+\Phi_2\not\!p\gamma_5+\Phi_3\not\!q\gamma_5+\Phi_4\gamma_5\Big),
\end{eqnarray}
where
\begin{eqnarray} \label{Pi1234}
\Phi_1&=&-g_{NN\pi}\lambda_N^2e^{-\frac{m_N^2}{M^2}-\frac{m_{N}^2}{M^{\prime^2}}}
+g_{NN^*\pi}\lambda_N\lambda_{N^*}e^{-\frac{m_N^2}{M^2}-\frac{m_{N^*}^2}{M^{\prime^2}}}
-g_{N^*N\pi}\lambda_N\lambda_{N^*}e^{-\frac{m_{N^*}^2}{M^2}-\frac{m_{N}^2}{M^{\prime^2}}}
\nonumber \\
&+&g_{N^*N^*\pi}\lambda_{N^*}^2e^{-\frac{m_{N^*}^2}{M^2}-\frac{m_{N^*}^2}{M^{\prime^2}}}
-g_{NN^{\prime}\pi}\lambda_{N} \lambda_{N^{\prime}}
e^{-\frac{m_{N}^2}{M^2}-\frac{m_{N^{\prime}}^2}{M^{\prime^2}}}
-g_{N^{\prime}N\pi}\lambda_{N}\lambda_{N^{\prime}}
e^{-\frac{m_{N^{\prime}}^2}{M^{2}}-\frac{m_{N}^2}{M^{\prime^2}}}
\nonumber \\
&-&g_{N^{\prime}N^{\prime}\pi}\lambda_{N^{\prime}}^2
e^{-\frac{m_{N^{\prime}}^2}{M^{2}}-\frac{m_{N^{\prime}}^2}{M^{\prime^2}}}+
g_{N^{\prime}N^{*}\pi}\lambda_{N^{\prime}}\lambda_{N^*}
e^{-\frac{m_{N^{\prime}}^2}{M^{2}}-\frac{m_{N^{*}}^2}{M^{\prime^2}}}-
g_{N^{*}N^{\prime}\pi}\lambda_{N^{\prime}}\lambda_{N^*}
e^{-\frac{m_{N^{*}}^2}{M^{2}}-\frac{m_{N^{\prime}}^2}{M^{\prime^2}}},
\nonumber \\
\Phi_2&=&-g_{NN^*\pi}\lambda_N\lambda_{N^*}(m_N+m_{N^*})e^{-\frac{m_N^2}{M^2}-\frac{m_{N^*}^2}
{M^{\prime^2}}}-g_{N^*N\pi}\lambda_N\lambda_{N^*}(m_N+m_{N^*})e^{-\frac{m_{N^*}^2}{M^2}-\frac{m_{N}^2}
{M^{\prime^2}}}
\nonumber \\
&-&g_{NN^{\prime}\pi}\lambda_N\lambda_{N^{\prime}}(m_{N^{\prime}}-m_{N})e^{-\frac{m_N^2}{M^2}
-\frac{m_{N^{\prime}}^2}
{M^{\prime^2}}}+g_{N^{\prime}N\pi}\lambda_N\lambda_{N^{\prime}}(m_{N^{\prime}}-m_{N})e^{-\frac{m_{N^{\prime}}^2}{M^2}
-\frac{m_N^2} {M^{\prime^2}}}
\nonumber \\
&-&
g_{N^{\prime}N^*\pi}\lambda_{N^{\prime}}\lambda_{N^*}(m_{N^{\prime}}+m_{N^*})
e^{-\frac{m_{N^{\prime}}^2}{M^2} -\frac{m_{N^{*}}^2}
{M^{\prime^2}}}-
g_{N^*N^{\prime}\pi}\lambda_{N^{\prime}}\lambda_{N^*}(m_{N^{\prime}}+m_{N^*})
e^{-\frac{m_{N^{*}}^2}{M^2} -\frac{m_{N^{\prime}}^2}
{M^{\prime^2}}},
\nonumber \\
\Phi_3&=&-g_{NN\pi}\lambda_N^2m_Ne^{-\frac{m_N^2}{M^2}-\frac{m_{N}^2}{M^{\prime^2}}}
+g_{NN^*\pi}\lambda_N\lambda_{N^*}m_Ne^{-\frac{m_N^2}{M^2}-\frac{m_{N^*}^2}{M^{\prime^2}}}
+g_{N^*N\pi}\lambda_N\lambda_{N^*}m_{N^*}e^{-\frac{m_{N^*}^2}{M^2}-\frac{m_{N}^2}{M^{\prime^2}}}
\nonumber \\
&-&g_{N^*N^*\pi}\lambda_{N^*}^2m_{N^*}e^{-\frac{m_{N^*}^2}{M^2}-\frac{m_{N^*}^2}{M^{\prime^2}}}-
g_{NN^{\prime}\pi}\lambda_{N}\lambda_{N^{\prime}}m_{N}e^{-\frac{m_{N}^2}{M^2}
-\frac{m_{N^{\prime}}^2}{M^{\prime^2}}}-g_{N^{\prime}N\pi}\lambda_{N}\lambda_{N^{\prime}}m_{N^{\prime}}
e^{-\frac{m_{N^{\prime}}^2}{M^2} -\frac{m_{N}^2}{M^{\prime^2}}}
\nonumber \\
&-&
g_{N^{\prime}N^{\prime}\pi}\lambda_{N^{\prime}}^2m_{N^{\prime}}e^{-\frac{m_{N^{\prime}}^2}{M^2}
-\frac{m_{N^{\prime}}^2}{M^{\prime^2}}}
+g_{N^{\prime}N^{*}\pi}\lambda_{N^{\prime}}\lambda_{N^*}m_{N^{\prime}}e^{-\frac{m_{N^{\prime}}^2}{M^2}
-\frac{m_{N^{*}}^2}{M^{\prime^2}}}+
g_{N^{*}N^{\prime}\pi}\lambda_{N^{\prime}}\lambda_{N^*}m_{N^{*}}e^{-\frac{m_{N^{*}}^2}{M^2}
-\frac{m_{N^{\prime}}^2}{M^{\prime^2}}} ,
\nonumber \\
\Phi_4&=&g_{NN\pi}\lambda_N^2q^2e^{-\frac{m_N^2}{M^2}-\frac{m_{N}^2}{M^{\prime^2}}}
-g_{N^*N\pi}\lambda_N\lambda_{N^*}(m_N^2+m_Nm_{N^*}-q^2)e^{-\frac{m_{N^*}^2}{M^2}-\frac{m_{N}^2}{M^{\prime^2}}}
\nonumber \\
&-&g_{N^*N^*\pi}\lambda_{N^*}^2q^2e^{-\frac{m_{N^*}^2}{M^2}-\frac{m_{N^*}^2}{M^{\prime^2}}}
+g_{NN^*\pi}\lambda_N\lambda_{N^*}(m_{N^*}^2+m_Nm_{N^*}-q^2)e^{-\frac{m_{N}^2}{M^2}
-\frac{m_{N^*}^2}{M^{\prime^2}}}
\nonumber \\
&+&g_{N^{\prime}N\pi}\lambda_{N}\lambda_{N^{\prime}}
(-m_N^2+m_Nm_{N^{\prime}}+q^2)e^{-\frac{m_{N^{\prime}}^2}{M^2}-\frac{m_{N}^2}{M^{\prime^2}}}
+g_{N^{\prime}N^{\prime}\pi}\lambda_{N^{\prime}}^2q^2e^{-\frac{m_{N^{^{\prime}}}^2}{M^2}-\frac{m_{N^{\prime}}^2}{M^{\prime^2}}}
\nonumber \\
&+&g_{NN^{\prime}\pi}\lambda_{N}\lambda_{N^{\prime}}
(-m_{N^{\prime}}^2+m_Nm_{N^{\prime}}+q^2)e^{-\frac{m_{N}^2}{M^2}-\frac{m_{N^{\prime}}^2}{M^{\prime^2}}}
-g_{N^*N^{\prime}\pi}\lambda_{N^{\prime}}\lambda_{N^*}
(m_{N^{\prime}}^2+m_{N^{\prime}}m_{N^{*}}-q^2)
\nonumber \\
&\times&
e^{-\frac{m_{N^*}^2}{M^2}-\frac{m_{N^{\prime}}^2}{M^{\prime^2}}}
+g_{N^{\prime}N^*\pi}\lambda_{N^{\prime}}\lambda_{N^*}
(m_{N^{*}}^2+m_{N^{\prime}}m_{N^{*}}-q^2)e^{-\frac{m_{N^{\prime}}^2}{M^2}-\frac{m_{N^{*}}^2}{M^{\prime^2}}}.
\end{eqnarray}
The $M^2$ and $M^{\prime^2}$ arising in these results are the Borel
mass parameters in the initial and final channels, respectively.

\subsection{OPE Side}

The OPE side of the correlation function is  calculated in deep Euclidean region. To this aim we use  the interpolating current
\begin{eqnarray}\label{InterpolatingCurrents}
J_{N^{('*)}}(y)=2\varepsilon_{ij\ell}\Big\{\Big(u^{i^T}(y)Cd^{j}(y)\Big)\gamma_{5}
u^{\ell}(y)+\beta\Big(u^{i^T}(y)C\gamma_5d^{j}(y)\Big)
u^{\ell}(y)\Big\},
\end{eqnarray}
which couples to the nucleon with both parities as well as
\begin{eqnarray}
J_{\pi}(0)=\frac{1}{\sqrt{2}}\Big(\bar{u}(0)i\gamma_{5}u(0)-\bar{d}(0)i\gamma_5d(0)\Big).
\end{eqnarray}
for the pion. In Eq.(\ref{InterpolatingCurrents}), $\beta$ is a general mixing parameter (with $\beta=-1$ corresponding to the Ioffe current) that we shall fix it by some physical considerations and $C$ is the charge conjugation operator.
 The
substitution of these interpolating currents in
Eq.~(\ref{CorrelationFunction}) is followed by possible contractions
of all quark pairs via Wick's theorem, and this leads  to
\begin{eqnarray}\label{correlfuncOPE}
\Pi^{OPE}(q)&=&2i^3\int d^{4}x\int d^{4}ye^{-ip\cdot
x}e^{ip^{\prime}\cdot y}\epsilon_{abc}\epsilon_{ij\ell}
\nonumber \\
&\times&
\Bigg\{-Tr\Big[S_d^{bj}(y-x)S_u^{\prime^{di}}(-x)\gamma_5S_u^{\prime^{ad}}(y)\Big]\Big(\gamma_5S_u^{c\ell}(y-x)\gamma_5\Big)
\nonumber \\
&+&Tr\Big[S_u^{\prime^{ai}}(y-x)S_d^{bj}(y-x)\Big]\Big(\gamma_5S_u^{cd}(y)\gamma_5S_u^{d\ell}(-x)\gamma_5\Big)
\nonumber \\
&-&\gamma_5S_u^{ci}(y-x)S_d^{\prime^{bj}}(y-x)S_u^{ad}(y)\gamma_5S_u^{d\ell}(-x)\gamma_5
\nonumber \\
&-&\gamma_5
S_u^{cd}(y)\gamma_5S_u^{di}(-x)S_d^{\prime^{bj}}(y-x)S_u^{a\ell}(y-x)\gamma_5
\nonumber \\
&-&Tr\Big[S_u^{\prime^{ai}}(y-x)S_d^{bd}(y)\gamma_5S_d^{dj}(-x)\Big]\Big(\gamma_5S_u^{c\ell}(y-x)\gamma_5\Big)
\nonumber \\
&-&\gamma_5S_u^{ci}(y-x)S_d^{\prime^{dj}}(-x)\gamma_5S_d^{\prime^{bd}}(y)S_u^{a\ell}(y-x)\gamma_5
\nonumber \\
&-&\beta\Bigg[Tr\Big[S_d^{bj}(y-x)\gamma_5S_u^{\prime^{di}}(-x)\gamma_5S_u^{\prime^{ad}}(y)\Big]\Big(\gamma_5S_u^{c\ell}(y-x)\Big)
\nonumber \\
&+&Tr\Big[\gamma_5S_d^{bj}(y-x)S_u^{\prime^{di}}(-x)\gamma_5S_u^{\prime^{ad}}(y)\Big]\Big(S_u^{c\ell}(y-x)\gamma_5\Big)
\nonumber \\
&-&Tr\Big[\gamma_5S_u^{\prime^{ai}}(y-x)S_d^{bj}(y-x)\Big]\Big(\gamma_5S_u^{cd}(y)\gamma_5S_u^{d\ell}(-x)\Big)
\nonumber \\
&-&Tr\Big[S_u^{\prime^{ai}}(y-x)\gamma_5S_d^{bj}(y-x)\Big]\Big(S_u^{cd}(y)\gamma_5S_u^{d\ell}(-x)\gamma_5\Big)
\nonumber \\
&+&\gamma_5S_u^{ci}(y-x)\gamma_5S_d^{\prime^{bj}}(y-x)S_u^{ad}(y)\gamma_5S_u^{d\ell}(-x)
\nonumber \\
&+&S_u^{ci}(y-x)S_d^{\prime^{bj}}(y-x)\gamma_5S_u^{ad}(y)\gamma_5S_u^{d\ell}(-x)\gamma_5
\nonumber \\
&+&S_u^{cd}(y)\gamma_5S_u^{di}(-x)S_d^{\prime^{bj}}(y-x)\gamma_5S_u^{a\ell}(y-x)\gamma_5
\nonumber \\
&+&\gamma_5S_u^{cd}(y)\gamma_5S_u^{di}(-x)\gamma_5S_d^{\prime^{bj}}(y-x)S_u^{a\ell}(y-x)
\nonumber \\
&+&Tr\Big[\gamma_5S_u^{\prime^{ai}}(y-x)S_d^{bd}(y)\gamma_5S_d^{dj}(-x)\Big]\Big(\gamma_5S_u^{c\ell}(y-x)\Big)
\nonumber \\
&+&Tr\Big[S_u^{\prime^{ai}}(y-x)\gamma_5S_d^{bd}(y)\gamma_5S_d^{dj}(-x)\Big]\Big(S_u^{c\ell}(y-x)\gamma_5\Big)
\nonumber \\
&+&S_u^{ci}(y-x)S_d^{\prime^{dj}}(-x)\gamma_5S_d^{\prime^{bd}}(y)\gamma_5S_u^{a\ell}(y-x)\gamma_5
\nonumber \\
&+&\gamma_5S_u^{ci}(y-x)\gamma_5S_d^{\prime^{dj}}(-x)\gamma_5S_d^{\prime^{bd}}(y)S_u^{a\ell}(y-x)
 \Bigg]
 \nonumber \\
&-&\beta^2\Bigg[Tr\Big[\gamma_5S_d^{bj}(y-x)\gamma_5S_u^{\prime^{di}}(-x)\gamma_5S_u^{\prime^{ad}}(y)\Big]S_u^{c\ell}(y-x)
\nonumber \\
&-&Tr\Big[\gamma_5S_u^{\prime^{ai}}(y-x)\gamma_5S_d^{bj}(y-x)\Big]\Big(S_u^{cd}(y)\gamma_5S_u^{d\ell}(-x)\Big)
\nonumber \\
&+&S_u^{ci}(y-x)\gamma_5S_d^{\prime^{bj}}(y-x)\gamma_5S_u^{ad}(y)\gamma_5S_u^{d\ell}(-x)
\nonumber \\
&+&S_u^{cd}(y)\gamma_5S_u^{di}(-x)\gamma_5S_d^{\prime^{bj}}(y-x)\gamma_5S_u^{a\ell}(y-x)
\nonumber \\
&+&Tr\Big[\gamma_5S_u^{\prime^{ai}}(y-x)\gamma_5S_d^{bd}(y)\gamma_5S_d^{dj}(-x)\Big]S_u^{c\ell}(y-x)
\nonumber \\
&+&S_u^{ci}(y-x)\gamma_5S_d^{\prime^{dj}}(-x)\gamma_5S_d^{\prime^{bd}}(y)\gamma_5S_u^{a\ell}(y-x)
 \Bigg]
 \Bigg\}~,
\end{eqnarray}
where $S^{\prime}=CS^TC$. In this equation, $S^{ij}_{q}(x)$
corresponds to the light quark propagator for which we use
\cite{Reinders}
\begin{eqnarray}\label{lightpropagator}
S_{q}^{ij}(x)&=& i\frac{\!\not\!{x}}{ 2\pi^2 x^4}\delta_{ij}
-\frac{m_q}{4\pi^2x^2}\delta_{ij}-\frac{\langle
\bar{q}q\rangle}{12}\Big(1 -i\frac{m_q}{4}
\!\not\!{x}\Big)\delta_{ij} -\frac{x^2}{192}m_0^2\langle
\bar{q}q\rangle\Big(1-i\frac{m_q}{6} \!\not\!{x}\Big)\delta_{ij}
\nonumber \\
&-&\frac{ig_s
G_{\theta\eta}^{ij}}{32\pi^2x^2}\big[\!\not\!{x}\sigma^{\theta\eta}+\sigma^{\theta\eta}\!\not\!{x}\big]
+\cdots \, ,
\end{eqnarray}
where $q$ is either $u$ or $d$ quark.
Here we shall remark that we consider the terms up to dimension five in OPE in the calculations. Note that in the expression of the quark propagator given in
 Eq. (\ref{lightpropagator}), the vacuum saturation is assumed for the quark fields. In this assumption, the diagrams with the quark-gluon mixed condensate
in which the gluon comes from a different quark propagator than the quark fields are missing. We separately calculate and add the contributions of such diagrams to the obtained expressions.
Figure 1 shows some typical diagrams that we take into account in the present study.
\begin{figure}[h!]
 \begin{center}
\includegraphics[width=15cm]{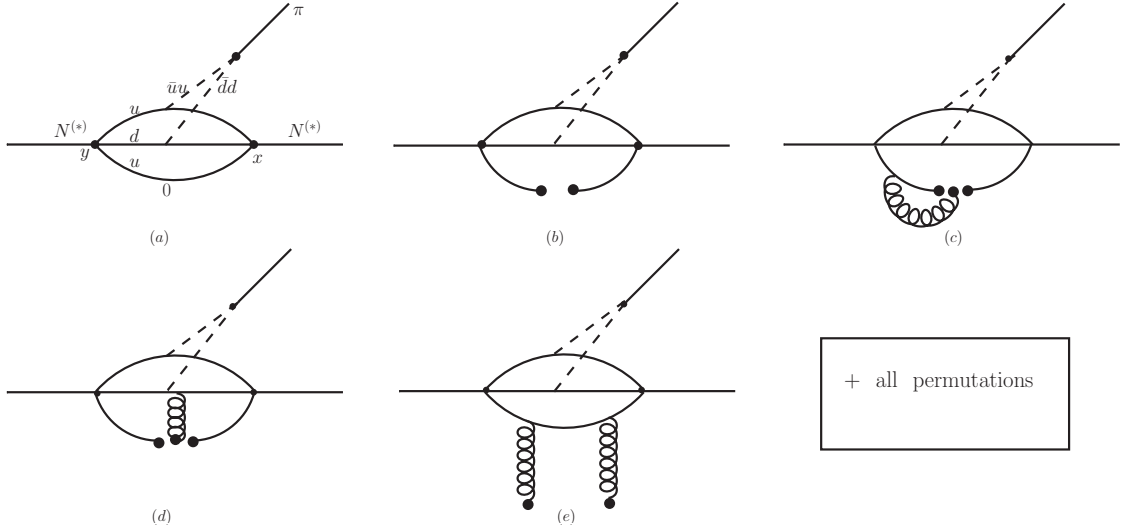}
\end{center} \caption{Typical diagrams taken into account in the calculations.}
\label{fig1a}
\end{figure}

 The OPE side in coordinate space  is obtained by inserting the above  propagator into
Eq.~(\ref{correlfuncOPE}). A Fourier transformation is
applied to transform the calculations to  the momentum space. To this end, the following expression  in $D$ dimension  is used:
\begin{eqnarray}\label{intyx}
\frac{1}{[A^2]^n}&=&\int\frac{d^Dt}{(2\pi)^D}e^{-it\cdot A}~i~(-1)^{n+1}~2^{D-2n}~\pi^{D/2}~
\frac{\Gamma(D/2-n)}{\Gamma(n)}\Big(-\frac{1}{t^2}\Big)^{D/2-n},
\end{eqnarray}
and  the four-integrals over $x$ and $y$ are performed after
the replacements $x_{\mu}\rightarrow i\frac{\partial}{\partial
p_{\mu}}$ and
 $y_{\mu}\rightarrow -i\frac{\partial}{\partial p'_{\mu}}$. After making use of the Feynman parametrization and
 the equation

\begin{eqnarray}\label{Int}
\int d^4t\frac{(t^2)^{\beta}}{(t^2+L)^{\alpha}}=\frac{i \pi^2
(-1)^{\beta-\alpha}\Gamma(\beta+2)\Gamma(\alpha-\beta-2)}{\Gamma(2)
\Gamma(\alpha)[-L]^{\alpha-\beta-2}},
\end{eqnarray}
 we get
\begin{eqnarray}\label{correlfuncOPELast}
\Pi^{OPE}(q)&=&\Pi_1(q^2)\not\!p\not\!q\gamma_5+\Pi_2(q^2)\not\!p\gamma_5+\Pi_3(q^2)\not\!q\gamma_5
+\Pi_4(q^2) \gamma_5,
\end{eqnarray}
where the functions $\Pi_i(q^2)$  include the
contributions coming from both the perturbative and non-perturbative
parts. These functions are given as
\begin{eqnarray}\label{QCDside1}
\Pi_i(q^2)=\int^{}_{}ds\int^{}_{}ds^{\prime}
\frac{\rho_i^{pert}(s,s^{\prime},q^2)+\rho_i^{non-pert}(s,s^{\prime},q^2)}{(s-p^2)
(s^{\prime}-p^{\prime^2})}~,
\end{eqnarray}
where the $\rho_i(s,s',q^2)$ appearing in this equation are corresponding to
spectral densities associated with different structures.   They are attained by  taking the imaginary parts of
the $\Pi_{i}$ functions, i.e.
$\rho_i(s,s',q^2)=\frac{1}{\pi}Im[\Pi_{i}]$. As examples, we present only the spectral densities
corresponding to the Dirac structure $\gamma_{5}$ here. They are obtained as
\begin{eqnarray}\label{rho1pert}
\rho_4^{pert}(s,s^{\prime},q^2)&=& \int_{0}^{1}dx \int_{0}^{1-x}dy
\frac{i^3}{128\pi^4y}(1-\beta)(7\beta+5)\Big[q^4x(x+y-1)\Big(1+x^2+x(y-1)
\nonumber \\
&-& y(1+y)\Big)
-q^2\Big(sxy(1-3x+3x^2+2xy-y^2)-s^{\prime}(x+y-1)(y-x-2xy
\nonumber \\
&+& 3x^2y-y^2+5xy^2)\Big)+2s^2x^2y^2
+ss^{\prime}y(x+4xy-5x^2y-5xy^2)+s^{\prime^2}y(x+y-1)
\nonumber \\
&\times& (2y^2+xy-y-1)\Big] \Theta\Big[L_1(s,s^{\prime},q^2)\Big]
,
\end{eqnarray}
and
\begin{eqnarray}\label{rho1nonpert}
\rho_4^{non-pert}(s,s^{\prime},q^2)&=&
\langle\alpha_s\frac{G^2}{\pi}\rangle\frac{i^3}{384\pi^2}
\Bigg\{\int_{0}^{1}dx \int_{0}^{1-x}dy
\frac{\beta-1}{y}\Big[(5+7\beta)(x^2-x+xy)-23y^2
\nonumber \\
&+&2y-2y\beta- 13y^2\beta)\Big]
 \Theta\Big[L_1(s,s^{\prime},q^2)\Big]+\frac{s^{\prime}(\beta-1)^2}{q^2}
 \Theta\Big[L_2(s,s^{\prime},q^2)\Big]
\nonumber \\
&+&\frac{s(\beta-1)^2}{q^2}\Theta\Big[L_3(s,s^{\prime},q^2)\Big]
 \Bigg\},
\end{eqnarray}
where
\begin{eqnarray}\label{teta1}
 L_1(s,s^{\prime},q^2)&=&q^2(1-x-y)+sxy+s^{\prime}y(1-x-y),
\nonumber \\
L_2(s,s^{\prime},q^2)&=&s^{\prime}
\nonumber \\
L_3(s,s^{\prime},q^2)&=&s~.
\end{eqnarray}
Here $\Theta[...]$ is the unit-step function.  For simplicity  we ignored to present  the terms containing the light quark mass in these formulas.

Now, we match the two sides in Borel scheme, after which we get four equations with nine couplings as unknowns. Hence, we need five more equations  to solve
nine equations with nine unknowns. We construct these extra equations by applying derivatives with respect to inverse of the Borel mass squares.  
As a result, we get the following expressions for the strong couplings that we are interested in their calculations in the present work:
\begin{eqnarray}\label{strongcouplingconstants}
g_{NN\pi}(q^2)&=&\frac{(m_u+m_d)(m_{\pi}^2-q^2)}{f_{\pi}\lambda_N^2m_{\pi}^2
(m_N+m_{N^*})^2(m_N-m_{N^{\prime}})^2}e^{\frac{m_N^2}{M^2}+\frac{m_{N}^2}{M^{\prime^2}}}
\left\{-m_{N^*}\frac{d(\widehat{\textbf{B}}\Pi_2)}{d\frac{1}{M^{\prime^2}}}\right.
\nonumber \\
&+&\left. \frac{d(\widehat{\textbf{B}}\Pi_4)}{d\frac{1}{M^{2}}}+m_N^2
\left[\frac{d(\widehat{\textbf{B}}\Pi_1)}{d\frac{1}{M^{2}}}-m_{N^*}\widehat{\textbf{B}}\Pi_3
+m_{N^{\prime}}(m_{N^*}\widehat{\textbf{B}}\Pi_1+\widehat{\textbf{B}}\Pi_3)\right]+m_{N^*}^2
\widehat{\textbf{B}}\Pi_4\right.
\nonumber \\
&+&\left. q^2
\frac{d(\widehat{\textbf{B}}\Pi_1)}{d\frac{1}{M^{2}}}+q^2m_{N^*}^2\widehat{\textbf{B}}\Pi_1-
m_{N^{\prime}}^2\left[m_{N^*}^2\widehat{\textbf{B}}\Pi_1+m_{N^*}\widehat{\textbf{B}}\Pi_3
-\widehat{\textbf{B}}\Pi_4-q^2\widehat{\textbf{B}}\Pi_1\right]
\right.
\nonumber \\
&-&\left.
m_N\left[-\frac{d(\widehat{\textbf{B}}\Pi_2)}{d\frac{1}{M^{2}}}+m_{N^*}^2(\widehat{\textbf{B}}\Pi_2+
\widehat{\textbf{B}}\Pi_3)+m_{N^{\prime}}^2(m_{N^*}\widehat{\textbf{B}}\Pi_1+
\widehat{\textbf{B}}\Pi_2+\widehat{\textbf{B}}\Pi_3)
\right.\right.
\nonumber \\
&+&\left.\left. m_{N^{\prime}}
\left(2\frac{d(\widehat{\textbf{B}}\Pi_1)}{d\frac{1}{M^{\prime^2}}}-\frac{d(\widehat{\textbf{B}}\Pi_1)}{d\frac{1}{M^{2}}}
-m_{N^*}^2\widehat{\textbf{B}}\Pi_1-3m_{N^*}\widehat{\textbf{B}}\Pi_2-2m_{N^*}\widehat{\textbf{B}}\Pi_3
+\widehat{\textbf{B}}\Pi_4 \right.\right.\right.
\nonumber \\
&+&\left.\left.\left.
q^2\widehat{\textbf{B}}\Pi_1\right)+m_{N^*}\left(\frac{d(\widehat{\textbf{B}}\Pi_1)}{d\frac{1}{M^{2}}}
-2\frac{d(\widehat{\textbf{B}}\Pi_1)}{d\frac{1}{M^{\prime^2}}}-\widehat{\textbf{B}}\Pi_4-q^2\widehat{\textbf{B}}\Pi_1
\right)
\right]+m_{N^{\prime}}\left[\frac{d(\widehat{\textbf{B}}\Pi_2)}{d\frac{1}{M^{\prime^2}}}
\right.\right.
\nonumber \\
&+&\left.\left. m_{N^*}
\left(-\frac{d(\widehat{\textbf{B}}\Pi_1)}{d\frac{1}{M^{2}}}+m_{N^*}\widehat{\textbf{B}}\Pi_3-\widehat{\textbf{B}}\Pi_4
-q^2\widehat{\textbf{B}}\Pi_1\right)\right]
 \right\},
 \nonumber \\
g_{N^*N\pi}(q^2)&=&\frac{(m_u+m_d)(m_{\pi}^2-q^2)}{f_{\pi}m_{\pi}^2\lambda_N\lambda_{N^*}
(m_N+m_{N^*})^2(m_N-m_{N^{\prime}})(m_{N^{\prime}}+m_{N^*})}e^{\frac{m_{N^*}^2}{M^2}+
\frac{m_{N}^2}{M^{\prime^2}}}
\nonumber \\
&\times&
 \left\{m_{N^\prime}\left(m_{N^*}\frac{d(\widehat{\textbf{B}}\Pi_1)}{d\frac{1}{M^{\prime^2}}}
 +\frac{d(\widehat{\textbf{B}}\Pi_2)}{d\frac{1}{M^{\prime^2}}}-m_{N^*}\frac{d(\widehat{\textbf{B}}\Pi_1)}{d\frac{1}{M^{2}}}\right)
 +\frac{d(\widehat{\textbf{B}}\Pi_4)}{d\frac{1}{M^{2}}}+m_{N}^2 m_{N^*}\widehat{\textbf{B}}\Pi_2\right.
\nonumber \\
&+&\left. m_{N}^2
\left(\frac{d(\widehat{\textbf{B}}\Pi_1)}{d\frac{1}{M^{2}}}
 -\frac{d(\widehat{\textbf{B}}\Pi_1)}{d\frac{1}{M^{\prime^2}}}-m_{N^\prime}\widehat{\textbf{B}}\Pi_2\right)
 +m_N\left[m_{N^*}\frac{d(\widehat{\textbf{B}}\Pi_1)}{d\frac{1}{M^{\prime^2}}}+\frac{d(\widehat{\textbf{B}}\Pi_2)}{d\frac{1}{M^{\prime^2}}}
\right.\right.
\nonumber \\
&-&\left.\left.
 m_{N^*}\frac{d(\widehat{\textbf{B}}\Pi_1)}{d\frac{1}{M^{2}}}+\frac{d(\widehat{\textbf{B}}\Pi_2)}{d\frac{1}{M^{2}}}
 +m_{N^\prime}\left(\frac{d(\widehat{\textbf{B}}\Pi_1)}{d\frac{1}{M^{2}}}-\frac{d(\widehat{\textbf{B}}\Pi_1)}{d\frac{1}{M^{\prime^2}}}
 +2m_{N^*}\widehat{\textbf{B}}\Pi_2\right)\right]
\right.
\nonumber \\
&+&\left.
 q^2\frac{d(\widehat{\textbf{B}}\Pi_1)}{d\frac{1}{M^{2}}}+m_{N^\prime}^2
 \left(m_{N^*}\widehat{\textbf{B}}\Pi_2+\widehat{\textbf{B}}\Pi_4+q^2\widehat{\textbf{B}}\Pi_1\right)
 \right\},
\nonumber \\
g_{NN^*\pi}(q^2)&=&\frac{(m_u+m_d)(m_{\pi}^2-q^2)}{f_{\pi}m_{\pi}^2\lambda_N\lambda_{N^*}
(m_N+m_{N^*})^2(m_N-m_{N^{\prime}})(m_{N^{\prime}}+m_{N^*})}e^{\frac{m_{N}^2}{M^2}+
\frac{m_{N^*}^2}{M^{\prime^2}}}
\nonumber \\
&\times&
 \left\{m_{N^*}^2\frac{d(\widehat{\textbf{B}}\Pi_1)}{d\frac{1}{M^{\prime^2}}}+
 m_{N^*}\frac{d(\widehat{\textbf{B}}\Pi_2)}{d\frac{1}{M^{\prime^2}}}-m_{N^*}^2
 \frac{d(\widehat{\textbf{B}}\Pi_1)}{d\frac{1}{M^{2}}}+m_{N^*}\frac{d(\widehat{\textbf{B}}\Pi_2)}{d\frac{1}{M^{2}}}-
 \frac{d(\widehat{\textbf{B}}\Pi_4)}{d\frac{1}{M^{2}}}
\right.
\nonumber \\
&+&\left.
 m_N(m_{N^\prime}-m_{N^*})\left(
 \frac{d(\widehat{\textbf{B}}\Pi_1)}{d\frac{1}{M^{\prime^2}}}-\frac{d(\widehat{\textbf{B}}\Pi_1)}{d\frac{1}{M^{2}}}+
 m_{N^\prime}\widehat{\textbf{B}}\Pi_2-m_{N^*}\widehat{\textbf{B}}\Pi_2\right)
\right.
\nonumber \\
& -&\left. m_{N^\prime}
\left(m_{N^*}\frac{d(\widehat{\textbf{B}}\Pi_1)}{d\frac{1}{M^{\prime^2}}}+
 \frac{d(\widehat{\textbf{B}}\Pi_2)}{d\frac{1}{M^{\prime^2}}}-m_{N^*}\frac{d(\widehat{\textbf{B}}\Pi_1)}{d\frac{1}{M^{2}}}
- m_{N^*}^2
 \widehat{\textbf{B}}\Pi_2\right)
\right.
\nonumber \\
&-&\left.
q^2\frac{d(\widehat{\textbf{B}}\Pi_1)}{d\frac{1}{M^{2}}}-m_{N^\prime}^2
 (\widehat{\textbf{B}}\Pi_4+q^2\widehat{\textbf{B}}\Pi_1)
 \right\},
\nonumber \\
g_{N^*N^*\pi}(q^2)&=&\frac{(m_u+m_d)(m_{\pi}^2-q^2)}{f_{\pi}m_{\pi}^2\lambda_{N^*}^2
(m_N+m_{N^*})^2(m_{N^*}+m_{N^{\prime}})^2}e^{\frac{m_{N^*}^2}{M^2}+
\frac{m_{N^*}^2}{M^{\prime^2}}}
\nonumber \\
&\times&
 \left\{-m_{N^*}^2\frac{d(\widehat{\textbf{B}}\Pi_1)}{d\frac{1}{M^{2}}}+m_{N^*}\frac{d(\widehat{\textbf{B}}\Pi_2)}{d\frac{1}{M^{2}}}
 -\frac{d(\widehat{\textbf{B}}\Pi_4)}{d\frac{1}{M^{2}}}-q^2\frac{d(\widehat{\textbf{B}}\Pi_1)}{d\frac{1}{M^{2}}}
 -m_{N^\prime}^2
 \widehat{\textbf{B}}\Pi_4\right.
\nonumber \\
&-&\left. m_{N^\prime}^2\left[
m_{N^*}\left(\widehat{\textbf{B}}\Pi_2+\widehat{\textbf{B}}\Pi_3\right)
 +q^2\widehat{\textbf{B}}\Pi_1\right]+m_N^2\left[m_{N^\prime}^2\widehat{\textbf{B}}\Pi_1+
 m_{N^\prime}\left(m_{N^*}\widehat{\textbf{B}}\Pi_1-\widehat{\textbf{B}}\Pi_3\right)
\right.\right.
\nonumber \\
&-&\left.\left. m_{N^*}
 \left(\widehat{\textbf{B}}\Pi_2+\widehat{\textbf{B}}\Pi_3\right)-\widehat{\textbf{B}}\Pi_4-q^2
 \widehat{\textbf{B}}\Pi_1\right]-m_{N^\prime}\left[\frac{d(\widehat{\textbf{B}}\Pi_2)}{d\frac{1}{M^{\prime^2}}}
 +m_{N^*}^2\widehat{\textbf{B}}\Pi_3+m_{N^*}
\right.\right.
\nonumber \\
&\times&\left.\left.
\left(2\frac{d(\widehat{\textbf{B}}\Pi_1)}{d\frac{1}{M^{\prime^2}}}
 -\frac{d(\widehat{\textbf{B}}\Pi_1)}{d\frac{1}{M^{2}}}+\widehat{\textbf{B}}\Pi_4+q^2\widehat{\textbf{B}}\Pi_1\right)\right]
+m_N\left[-\frac{d(\widehat{\textbf{B}}\Pi_2)}{d\frac{1}{M^{\prime^2}}}+m_{N^*}^2m_{N^\prime}
\widehat{\textbf{B}}\Pi_1 \right.\right.
\nonumber \\
&-&\left.\left.
m_{N^*}^2\widehat{\textbf{B}}\Pi_3-m_{N^\prime}\left(
\frac{d(\widehat{\textbf{B}}\Pi_1)}{d\frac{1}{M^{2}}}+m_{N^\prime}\widehat{\textbf{B}}\Pi_3+\widehat{\textbf{B}}\Pi_4
+q^2\widehat{\textbf{B}}\Pi_1 \right)+m_{N^*}
\frac{d(\widehat{\textbf{B}}\Pi_1)}{d\frac{1}{M^{2}}} \right.\right.
\nonumber \\
&-&\left.\left.
m_{N^*}\left(2\frac{d(\widehat{\textbf{B}}\Pi_1)}{d\frac{1}{M^{\prime^2}}}
-m_{N^\prime}^2\widehat{\textbf{B}}\Pi_1+3m_{N^\prime}\widehat{\textbf{B}}\Pi_2+2m_{N^\prime}
\widehat{\textbf{B}}\Pi_3+\widehat{\textbf{B}}\Pi_4+q^2\widehat{\textbf{B}}\Pi_1\right)\right]
 \right\},
\nonumber \\
\end{eqnarray}
where $\widehat{\textbf{B}}\Pi_i$ show the Borel transformed form of $\Pi_i$ functions and $\frac{d(\widehat{\textbf{B}}\Pi_i)}{d\frac{1}{M^{(')2}}}$ are their derivatives with
 respect to inverse of the Borel mass squares. We also apply the continuum subtraction in the initial and final channels in this step and this add two more auxiliary parameters
$s_0$ and $s'_0$ that they should also be fixed.

\section{Numerical results}

 This section contains the numerical analysis and  our discussion on the dependence of the results  on $Q^2=-q^2$.
 The numerical analysis requires some input parameters given in table 1. From the sum rules for the coupling constants it is also clear that we need to know the residues of $N$ and $N^*$ baryons. We use their
$\beta$-dependent expressions calculated in Ref. \cite{Azizi3} adding also the contribution of  $N (1440)$ to the mass sum rules.
\begin{table}[ht]\label{Table1}
\centering \rowcolors{1}{lightgray}{white}
\begin{tabular}{cc}
\hline \hline
   Parameters  &  Values
           \\
\hline \hline $ \langle\frac{\alpha_sG^2}{\pi}\rangle $       &
$(0.012\pm0.004)$ $~\mbox{GeV$^4$}$
\cite{belyaev}   \\
$m_{d}$              & $4.8^{+0.5}_{-0.3}~\mbox{MeV}$\cite{Olive}\\
$ m_{u} $            &$2.3^{+0.7}_{-0.5}~\mbox{MeV}$ \cite{Olive}\\
$ m_{N} $      &   $ (938.272046\pm0.000021)~\mbox{MeV}$  \cite{Olive} \\
$ m_{N^*} $      &   $ 1525~TO~1535~\mbox{MeV}$  \cite{Olive} \\
$ m_{\pi} $      &   $ (134.9766\pm0.0006) ~\mbox{MeV} $ \cite{Olive}  \\
$ f_{\pi} $      &   $(130.41\pm0.03\pm0.20) ~\mbox{MeV}$
\cite{Rosner} \\
 \hline \hline
\end{tabular}
\caption{Input parameters used in the calculations.}
\end{table}

In addition to the input parameters given in the table 1 there are
four  parameters, $M^2$, $M'^2$, $s_0$ and $s'_0$ that should be fixed. By virtue of being auxiliary parameters, the results
should practically be independent of them as much as possible. This necessitates
a search for  working regions of these parameters. Considering the relations
with the first excited states in the initial and final channels, that
is,  the energy that characterizes the beginning of the
continuum; as well as the fact that the sum rules obtained contain all three states in the physical sides,  we determine that the suitable interval for both continuum 
thresholds is $(2.64-2.74)~GeV^2$. As for the
Borel mass parameters, the analysis are done over the criteria that
the contributions of the higher states and continuum are
sufficiently suppressed and the contributions of the operators with
higher dimensions are small. Our analysis based on these
criteria leads to the intervals $(1.0-3.0)~GeV^2$ and $(2.0-4.0)~GeV^2$ for the Borel parameters in the $N$ and $N^*$ channels, respectively.

As already mentioned, in our calculations, we use the most general form
of the interpolating field for nucleon which is composed of two independent
interpolating fields connected by a mixing parameter $\beta$. In the
analysis, we need the values of this auxiliary  parameter for which we have
weak dependence of the results on this parameter. Our numerical analysis shows that in the intervals  $-1\leq cos \theta\leq-0.5$ and
$0.5\leq cos \theta\leq1$, common for all vertices, the results depend weakly on this parameter. Note that we use $\cos\theta$ where $\theta=\tan^{-1}(\beta)$
 to explore the whole range ($-\infty$, $+\infty$) for $\beta$ via $-1\leq\cos\theta\leq1$.
Considering the maximum  contribution of the ground state pole to
the sum rules, we found that $\cos\theta=-0.6$, common for all vertices, is roughly the optimum value that leads to the largest $pole/total~contribution$ ratio. We  use this value to extract the values of
the coupling constants under consideration. We shall note that by the above working regions for the continuum thresholds and  Borel parameters as well as  the value of $\cos\theta$, not only the pole
contribution is maximum ($\simeq 70\%$ of the total contribution), but also the series of sum rules properly converge, i.e., the perturbative part constitutes roughly $78\%$  and
the term with higher dimension constitutes less than $5\%$ of the pole contribution.

%
%

%
%

\begin{figure}[h!]
\includegraphics[width=8cm]{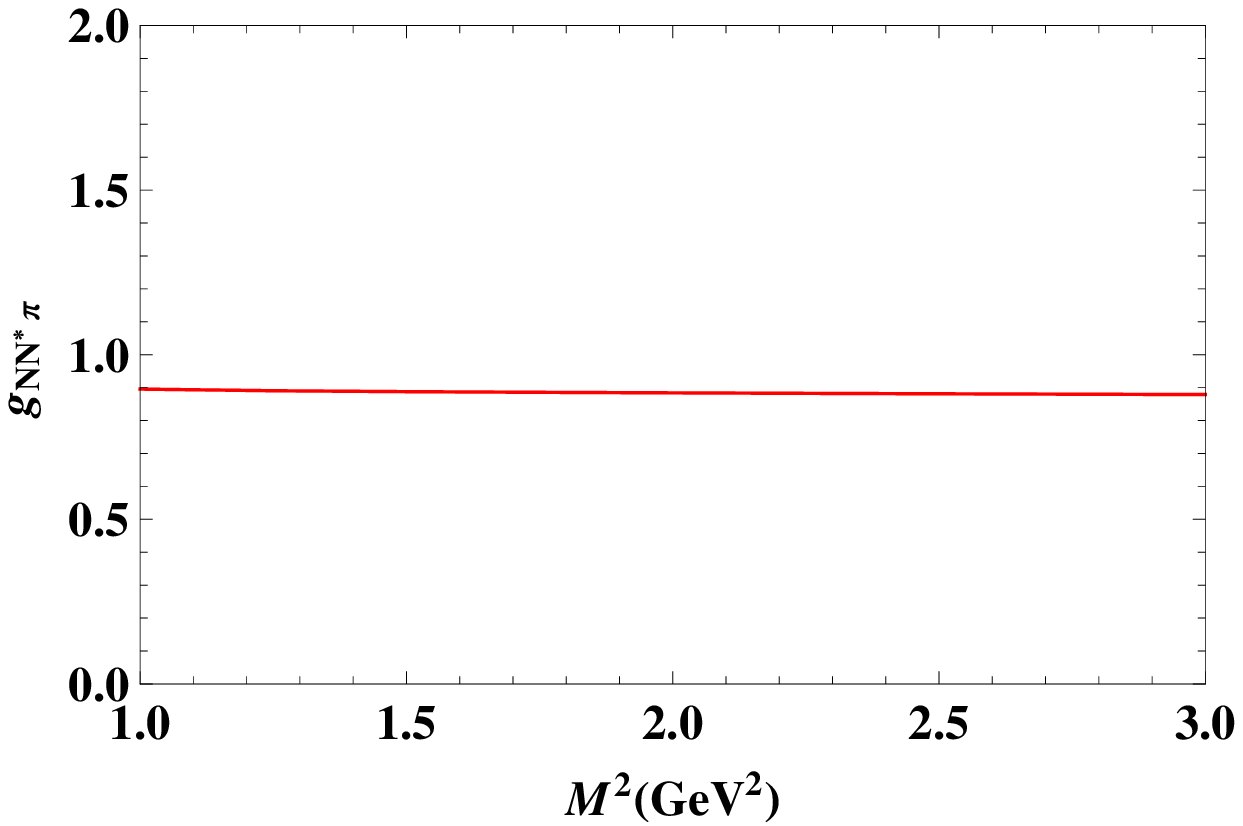}
\includegraphics[width=8cm]{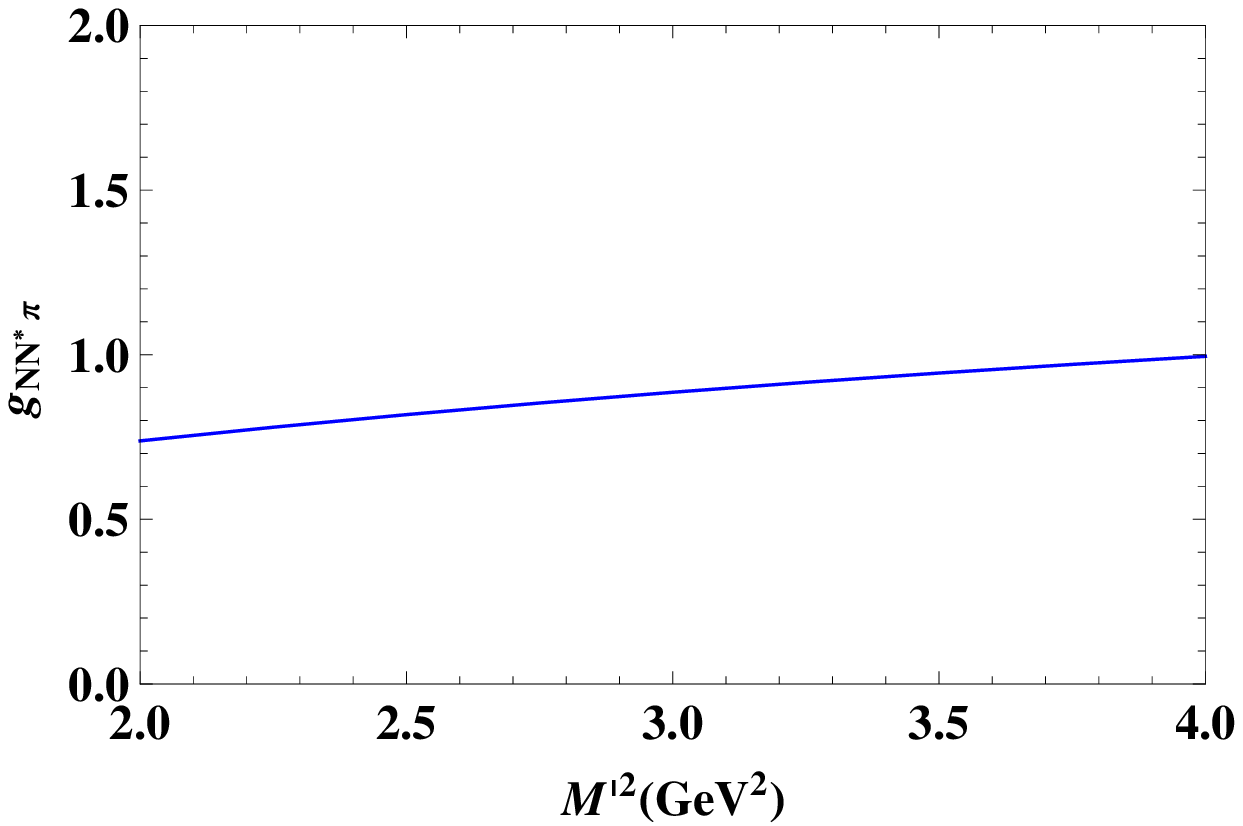}
\caption{\textbf{Left:} $g_{NN^*\pi}$ as a function of $M^2$ at
average values of continuum thresholds, $\cos\theta=-0.6$,
$Q^2=1GeV^2$ and $M^{'2}=3GeV^2$. \textbf{Right:} $g_{NN^*\pi}$ as
a function of $M^{'2}$ at average values of continuum thresholds,
$\cos\theta=-0.6$, $Q^2=1GeV^2$ and $M^{2}=2GeV^2$ }
\label{figborel}
\end{figure}

In this part we would like to show how the results of strong coupling constants  depend on the Borel parameters. To this end, we plot the dependence of $g_{NN^*\pi}$, as an example,
on these parameters at the average
values of the continuum thresholds, $\cos\theta=-0.6$ and $Q^2=1GeV^2$ in figure \ref{figborel}. From this figure we see that the results weakly depend on the Borel parameters in their working interval.

\begin{figure}[h!]
\includegraphics[width=8cm]{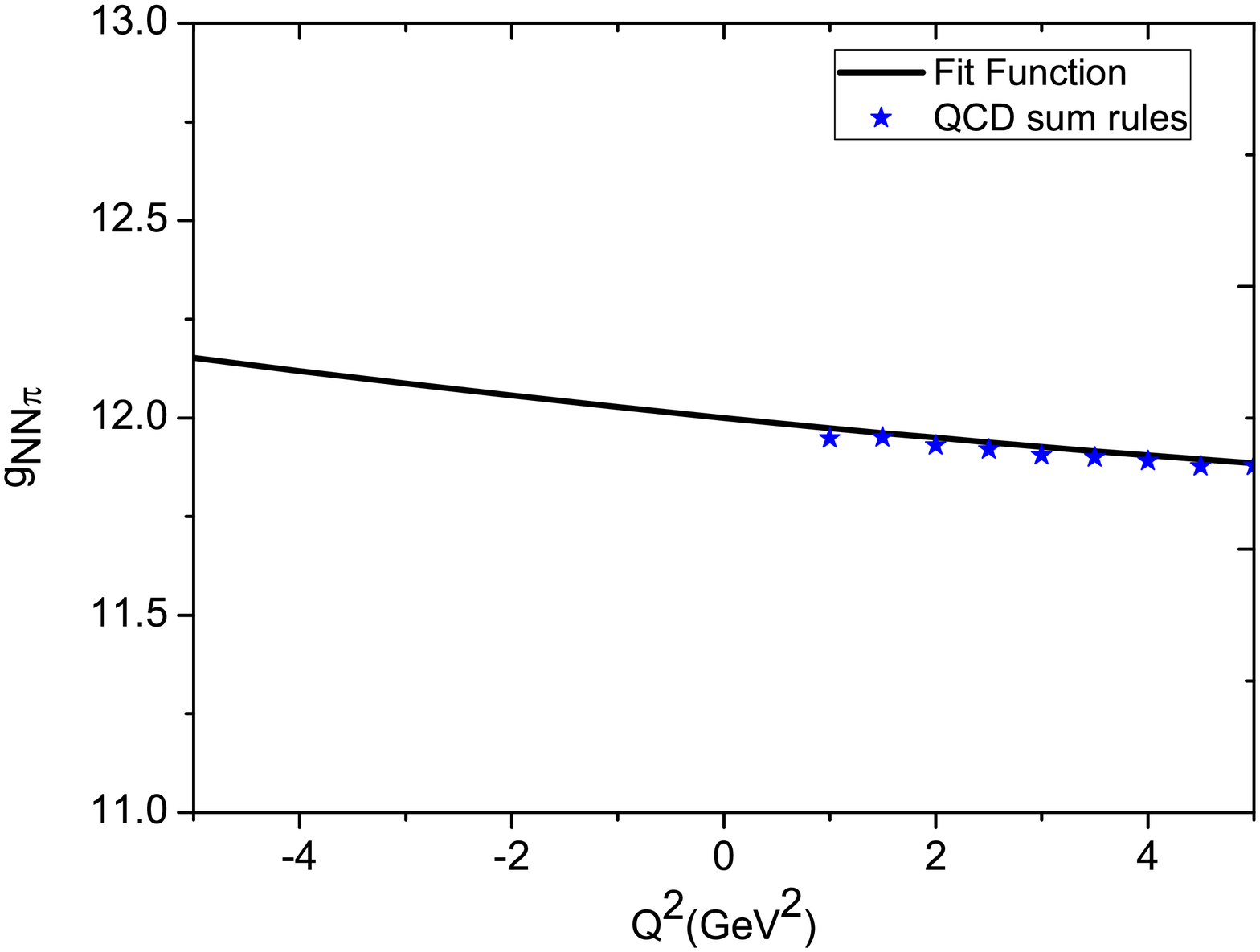}
\includegraphics[width=8cm]{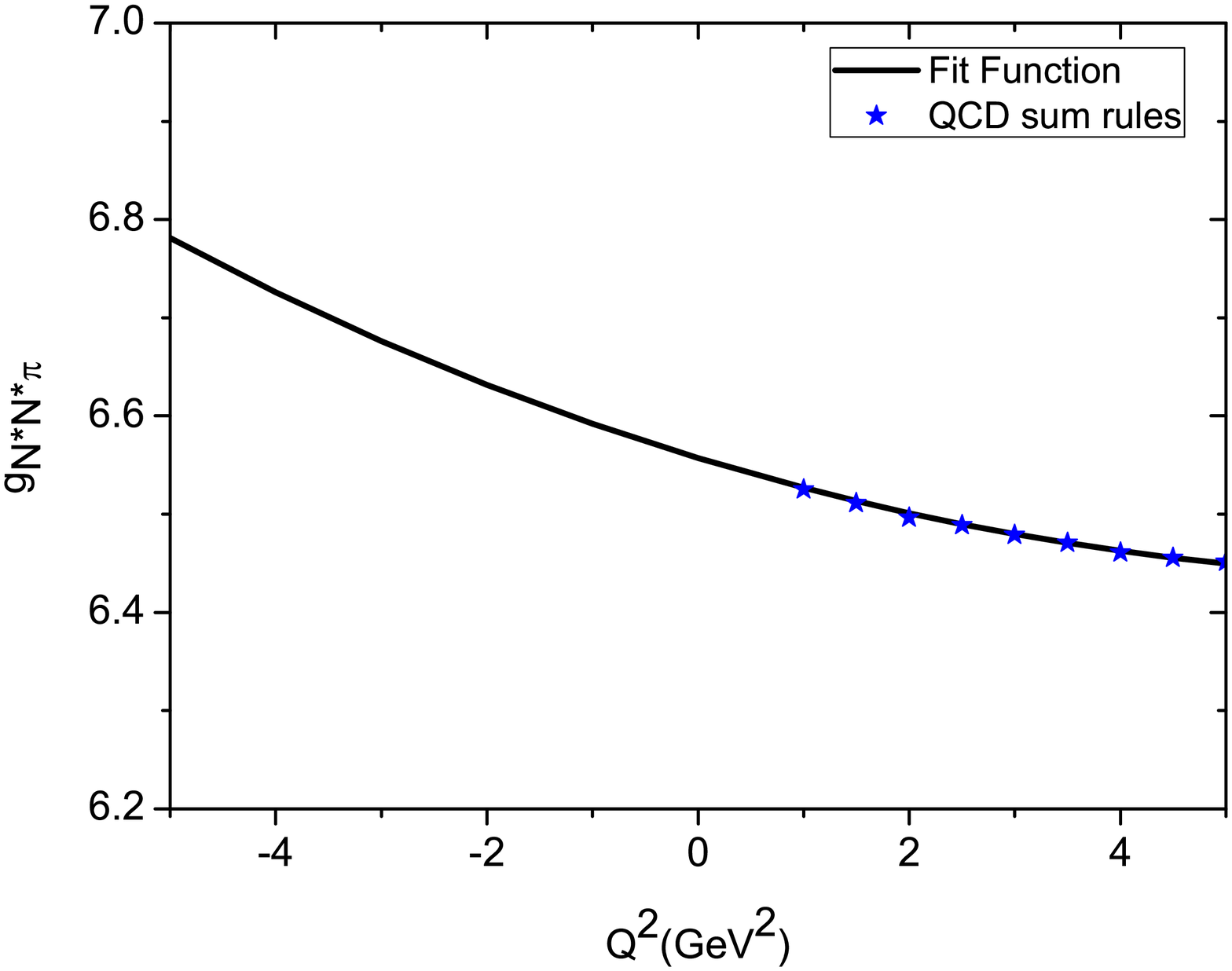}
\caption{\textbf{Left:} $g_{NN\pi}$ as a function of $Q^2$.
\textbf{Right:} $g_{N^*N^*\pi}$ as a function of $Q^2$ }\label{fig3}
\label{fig1a}
\end{figure}

\begin{figure}[h!]
\includegraphics[width=8cm]{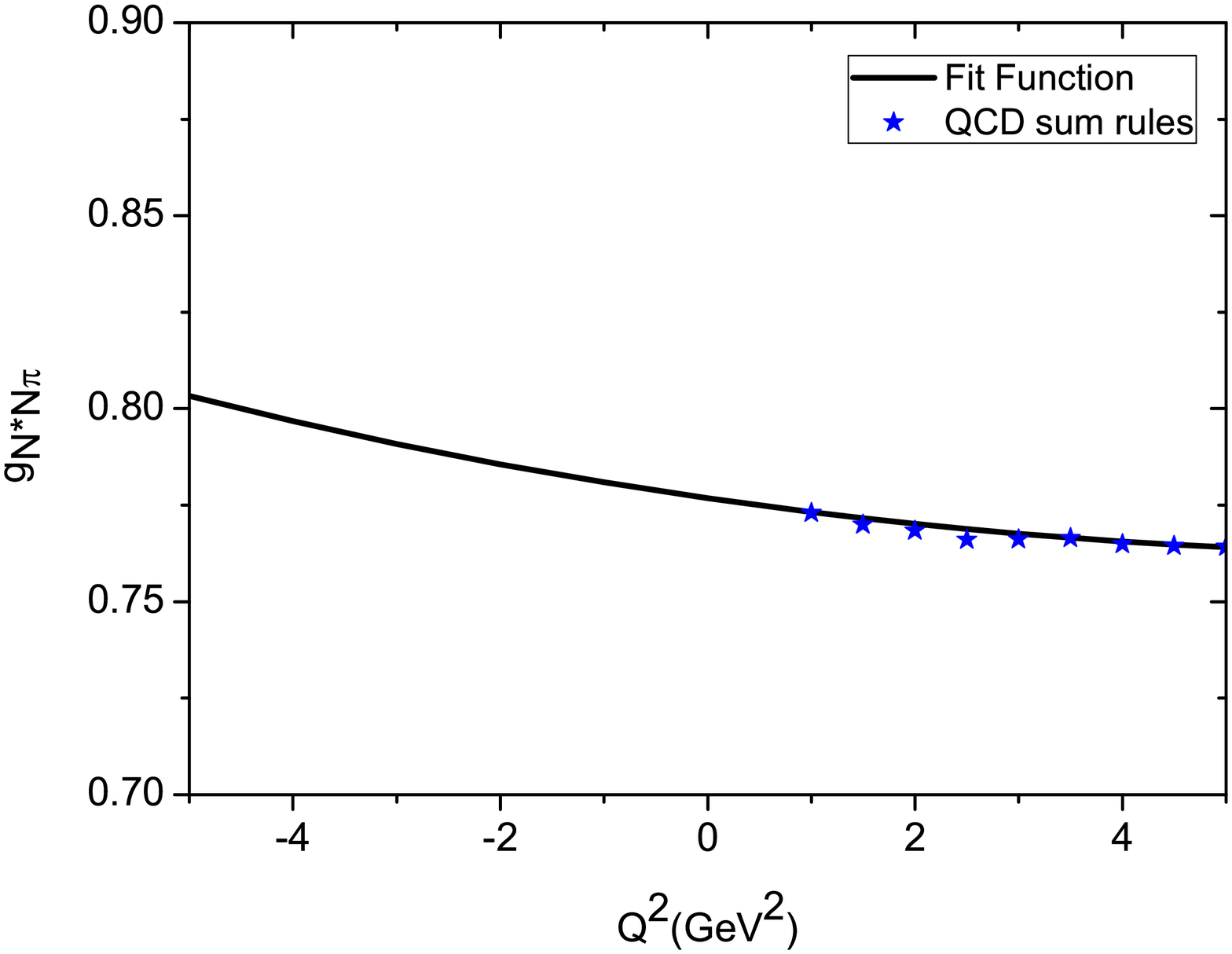}
\includegraphics[width=8cm]{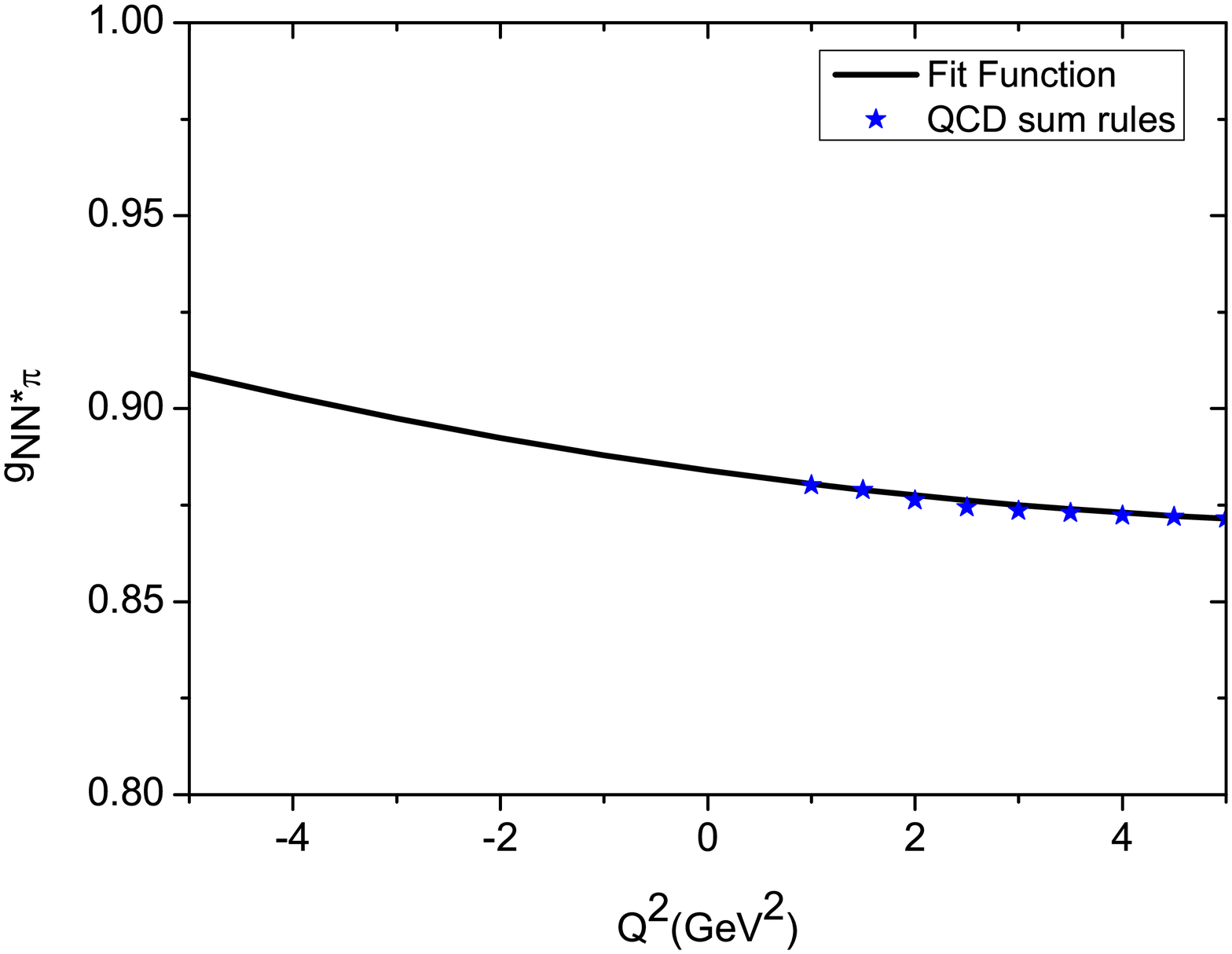}
\caption{\textbf{Left:} $g_{N^*N\pi}$ as a function of $Q^2$.
\textbf{Right:} $g_{NN^*\pi}$ as a function of $Q^2$ }
\label{fig4}
\end{figure}

Now, we proceed to discuss the behavior of the coupling constants with respect to
$Q^2$. Our calculations show that the
following fit function  describes well the strong couplings under consideration:
\begin{eqnarray}\label{fitfunc}
g_{N^{(*)}N^{(*)}\pi}(Q^2)=\frac{f_0}{1-a\frac{Q^2}{m_{N^{(*)}}^2}+b\Big(\frac{Q^2}{m_{N^{(*)}}^2}\Big)^2}.
\end{eqnarray}
Table \ref{fitparam} presents the values of fit parameters, $f_0$, $a$ and
$b$, for each coupling form factor. Figures (\ref{fig3}) and (\ref{fig4}) show the dependence of the strong  coupling constants under consideration on $Q^2$ for both sum rules and fit
results. From these figures, we observe that the above fit function  reproduces well the QCD sum rules results up to the truncated points.


 The
usage of the fit function at $Q^2=-m_{\pi}^2$ leads us to the
value of strong coupling constant for each considered transition as presented  in table \ref{couplingconstant} with the errors arising
from the uncertainties of the input parameters as well as
those coming from the determination of the working regions
of the auxiliary parameters.
 From this table we see that the couplings associated to the $N^*N\pi$ and  $NN^*\pi$ vertices are strongly suppressed compared to
 those related to two other
 vertices. The strong coupling   $g_{N^*N^*\pi}$ is obtained to be equal to roughly half of that of  $g_{NN\pi}$. From our results we also see
that the results for $g_{N^*N\pi}$ and $g_{NN^*\pi}$ are very
close to each other, which is an expected situation.
The result obtained for $g_{NN\pi}$
  is, within the errors, in good agreement with the results of
Refs.~\cite{Birse,Birse1,Kim1,Kim2} that obtain $g_{NN\pi } = 12 \pm
5$~\cite{Birse,Birse1}, $g_{NN\pi } = 9.76 \pm 2.04$~\cite{Kim1} and
$g_{NN\pi } = 13.3\pm 1.2$~\cite{Kim2}. Our prediction on the  $g_{NN^*\pi}$ is also consistent with the
 results of \cite{Hosaka,Zhu} that extract the value ~ $g_{ N
N^*\pi}\sim 0.7$ from the experimental data, but it differs considerably from the result of \cite{Zhu} which obtains $g_{
N N^*\pi} = (-)(0.08 \pm 0.06)$
from light cone QCD sum rules.
Our result on the
strong coupling constant $g_{N^*N^*\pi}$  can be checked via different phenomenological approaches as well as in future experiments.
\begin{table}[h]
\renewcommand{\arraystretch}{1.5}
\addtolength{\arraycolsep}{3pt}
$$
\begin{array}{|c|c|c|c|c|c|}
\hline \hline
       \mbox{}    & f_0 & a &b      \\
\hline
  \mbox{$g_{NN\pi}$} &12.001\pm 3.480&-0.005\pm0.001&(-3.143\pm0.943)10^{-4} \\
  \hline
  \mbox{$g_{N^*N^*\pi}$} &6.557\pm 1.961&-0.012\pm0.003&(-1.821\pm0.546)10^{-3} \\
  \hline
   \mbox{$g_{N^*N\pi}$} &0.777\pm 0.233&-0.012\pm0.003&(-1.820\pm0.546)10^{-3} \\
  \hline
  \mbox{$g_{NN^*\pi}$} &0.884\pm 0.256&-0.004\pm0.001&(-2.070\pm0.621)10^{-4}\\
                        \hline \hline
\end{array}
$$
\caption{Parameters appearing in the fit function.}
\label{fitparam}
\renewcommand{\arraystretch}{1}
\addtolength{\arraycolsep}{-1.0pt}
\end{table}
\begin{table}[h]
\renewcommand{\arraystretch}{1.5}
\addtolength{\arraycolsep}{3pt}
$$
\begin{array}{|c|c|c|c|c|}
\hline \hline
       \mbox{$g_{NN\pi}$}& \mbox{$g_{N^*N^*\pi}$} &\mbox{$g_{N^*N\pi}$}&\mbox{$g_{NN^*\pi}$}      \\
\hline
  12.012\pm3.608 &6.564\pm 1.842&0.782\pm0.233&0.882\pm0.264 \\
                         \hline \hline
\end{array}
$$
\caption{Values of the strong coupling constants.}
\label{couplingconstant}
\renewcommand{\arraystretch}{1}
\addtolength{\arraycolsep}{-1.0pt}
\end{table}

To sum up, we have calculated the   couplings
 $g_{NN\pi}$, $g_{N^*N\pi}$, $g_{NN^*\pi}$ and
$g_{N^*N^*\pi}$ using three point QCD sum rules. We used the most general form of the interpolating current for the nucleon. After fixing the
auxiliary parameters entering the calculations, we extracted the values of those couplings. We have found that the couplings $g_{N^*N\pi}$ and  $g_{NN^*\pi}$ are strongly suppressed.
The value of the coupling constant $g_{N^*N^*\pi}$ is also obtained to be roughly half of $g_{NN\pi}$. Our results on $g_{NN\pi}$ and  $g_{NN^*\pi}$ are in agreement with those of
\cite{Birse,Birse1,Kim1,Kim2,Zhu} which extract the results from different models as well as from the experimental data on the decay width of the corresponding transitions. Our result on $g_{NN^*\pi}$
considerably differs from the result of light cone QCD sum rules obtained in  \cite{Zhu}. This inconsistency can be attributed to the fact that the masses of the nucleon $N$ and the negative parity $N^*$
differ considerably and usage of the light cone QCD sum rules, as is done in \cite{Zhu}, for such vertex is problematic. Our prediction on the strong coupling constant $g_{N^*N^*\pi}$  can be checked
 via different phenomenological approaches as well as in future experiments.


\end{document}